\documentclass[A4,11pt,epsf,oneside]{article}
\usepackage{amssymb, amscd, amsmath,amsthm,amsfonts}
\usepackage{epsfig}
\usepackage{color}
\usepackage{capt-of}
\usepackage{float}

\begin{document}
\renewcommand{\thefootnote} {\fnsymbol{footnote}}
\setcounter{page}{1}

\title{\textbf{Approximate Noether Symmetries from Lagrangian for
Plane Symmetric Spacetimes}}
\author{\textbf{Farhad Ali and Tooba Feroze}
 \vspace{1cm}\\
{\normalsize \it \textbf{School of Natural Sciences, National}}\\
{\normalsize \it \textbf{University of Sciences and Technology, Islamabad,}}
       {\normalsize \textbf{\emph{Pakistan}} {\tt\textbf{}}}}
       \baselineskip=5mm
\baselineskip=7mm
 \maketitle
\begin{abstract}
Noether symmetries from geodetic Lagrangian for time-conformal plane symmetric spacetime are presented.
Here, time conformal factor is used to find the approximate Noether symmetries. This is a generalization of the idea discussed by I. Hussain and A. Qadir [3,4], where they obtained approximate Noether symmetries
from Lagrangian for a particular plane symmetric static spacetime. In the present article, the most general plane
symmetric static spacetime is considered and perturb it by introducing a general time
conformal factor $e^{\epsilon f(t)}$, where $\epsilon$ is very small which causes the
perturbation in the spacetime. Taking the perturbation up to the first order,
we find all Lagrangian for plane symmetric spactimes from which approximate Noether symmetries exist.
\end{abstract}
\section{Introduction}
Energy and momentum are important quantities whose definition have been a focus of many investigations in general relativity. Unfortunately, there is still no generally accepted definition of energy and momentum in general relativity. Therefore, different attempts are made to define these quantities.\\ A. Einstein  and N. Rosen [1] in their paper ``On gravitational waves" showed that the gravitational wave imparts energy from the source, the ultimate question arises how to find that energy which imparted by the gravitational wave from the source while in one of his paper A. H. Taub [2], proved that ``A spacetime with plane symmetry with $R_{\mu\nu}=0$ admits a coordinate system where the line element is independent of time \emph{t}, that is static", this indicates that we can not find out an exactly plane symmetric gravitational wave spacetime. So we can only find plane symmetric static vacuum solution(s) of the Einstein field equation\cite{Fa}. Now if there is no exact plane symmetric gravitational wave spacetime then how can one find out the energy content of the gravitational wave spacetime.\\ Recently, approximate Noether symmetries are used to define energy in gravitational wave like spacetime by considering a particular spacetime [3,4].\\ We used this technique of approximate Noether's symmetries to find out plane symmetric non-static spacetimes which are not exactly gravitational wave spacetimes resulting in approximate gravitational wave spacetimes for which $R_{\mu\nu}\rightarrow 0$ as $x\rightarrow\infty$.\\    Here, we find out the approximate Noether symmetries from the Lagrangian of the general plane symmetric first order perturbed spacetime for the investigation of energy and momentum. By considering plane symmetric static spacetimes as they play an important role in the study of gravitational waves whose static part ensures the conservation of energy. As waves cannot be static, therefore, by introducing a general time-conformal factor $\epsilon f(t)$ where $\epsilon$ is a  small arbitrary parameter. Taylor's expansion then leads to the first order approximate (perturbed) plane symmetric metric and the corresponding approximate Lagrangian. This approximate Lagrangian is then used to find the approximate Noether symmetries. We get many cases in which exact symmetries for the static spacetimes along with approximate part  appear. Among them time like infinitesimal generator $\frac{\partial}{\partial t}$ is the main focus of this paper which corresponds to the energy content in the respective spacetimes.   \\
For the general plane symmetric static spacetime [5]
\begin{equation}
ds^{2}_e=e^{\nu(x)}dt^{2}-dx^{2}-e^{\mu(x)}({dy}^{2}+{dz}^{2}),
\end{equation}
the corresponding Lagrangian is
\begin{equation}
L_e=e^{\nu(x)}\dot{t}^{2}-\dot{x}^{2}-e^{\mu(x)}(\dot{y}^{2}+\dot{z}^{2}).
\end{equation}
Now, introducing the general time conformal factor $f(t)$ we have
\begin{equation}ds^{2}=e^{\epsilon f(t)}ds^{2}_e,
\end{equation} and the corresponding perturbed Lagrangian is
\begin{equation}
L=e^{\epsilon f(t)}L_e.
\end{equation}
The first order perturbed metric and Lagrangian are
\begin{eqnarray}
ds^{2}&=&e^{\nu(x)}dt^{2}-dx^{2}-e^{\mu(x)}({dy}^{2}+{dz}^{2})+\epsilon
f(t)\{e^{\nu(x)}{dt}^{2}-dx^2-e^{\mu(x)}({dy}^{2}+{dz}^{2})\}, \label{4}\\
L&=&e^{\nu(x)}\dot{t}^{2}-\dot{x}^{2}-e^{\mu(x)}(\dot{y}^{2}+\dot{z}^{2})+\epsilon
f(t)\{e^{\nu(x)}\dot{t}^{2}-\dot{x}^2-e^{\mu(x)}(\dot{y}^{2}+\dot{z}^{2})\}. \label{4a}
 \end{eqnarray}
or
\begin{equation}
L=L_e+\epsilon L_a,
\end{equation}
where $L_e$ and $L_a$ are the exact and the perturbed parts respectively. This formation will give approximate part in the symmetry that corresponds the conservation of energy of the exact spacetime. This symmetry may be used for defining energy in gravitational waves.

\section{Approximate Noether symmetry}
An operator, $\textbf{X}$  is said to be an
approximate Noether symmetry generator if it satisfies [5]
\begin{equation}
\mathbf{X}^1L+(D\xi)L=DA,  \label{1}
\end{equation}
where $\mathbf{X}^1$ is the first order prolongation of the first
order approximate Noether symmetry $\textbf{X}=\mathbf{X}_e+\epsilon \mathbf{X}_a$.
Here, \begin{equation}
\textbf{X}_e=\xi_e\frac{\partial}{\partial
s}+\eta^{0}_e\frac{\partial}{\partial
t}+\eta^{1}_e\frac{\partial}{\partial
x}+\eta^{2}_e\frac{\partial}{\partial
y}+\eta^{3}_e\frac{\partial}{\partial z},
\end{equation}
and
\begin{equation}
\mathbf{X}_a=\xi_a\frac{\partial}{\partial
s}+\eta^{0}_a\frac{\partial}{\partial
t}+\eta^{1}_a\frac{\partial}{\partial
x}+\eta^{2}_a\frac{\partial}{\partial
y}+\eta^{3}_a\frac{\partial}{\partial z},
\end{equation} are respectively the exact and the approximate parts of $\mathbf{X}$.
In eq. $(\ref{1})$, $D$ is differential operator of the form
\begin{equation}D=\frac{\partial}{\partial
s}+\dot{t}\frac{\partial}{\partial
t}+\dot{x}\frac{\partial}{\partial
x}+\dot{y}\frac{\partial}{\partial
y}+\dot{z}\frac{\partial}{\partial z},
\end{equation}and $A=A_e+\epsilon A_a$ is gauge function whose exact and the approximate parts are $A_e$ and $A_a$.\\
The eq. $(\ref{1})$ splits into two parts as [7]
\begin{eqnarray}
\mathbf{X}_eL_e+(D\xi_e)L_e=DA_e, \label{2}\\
\textbf{X}^{1}_aL_e+\textbf{X}^{1}_eL_a+
(D\xi_e)L_a+(D\xi_a)L_e=DA_a. \label{3}
\end{eqnarray}
Here, $\eta^{i}_e,\eta^{i}_a,\xi_e,\xi_a ,A_e$, and $A_a$
are functions of $s,t,x,y,z$ and $\dot{\eta^{i}_e}, \dot{\eta^{i}_a}$ are functions of $s,t,x,y,z,$ $\dot{t},\dot{x},\dot{y},\dot{z}$ and ``$\ \dot{}\ $" denotes
differentiation with respect to $s$. Solutions of eq. $(\ref{2})$ are already known [8]. In this paper, we are interested in finding the solutions of eq. $(\ref{3})$ as these solutions will give us all those plane symmetric spacetimes whose Lagrangian may have approximate Noether symmetry(ies). Using the expressions of prolongation coefficients in eq. $(\ref{3})$ we obtain a system of Noether symmetries governing partial differential equations whose solutions provide the required results.
\section{System of Noether Symmetries Governing partial differential equations}
From eq. (8) we obtaine the following system of 19 partial differential equations:
\begin{equation}
\xi_{a,t}=0 ,\xi_{a,x}=0
,\xi_{a,y}=0 ,\xi_{a,z}=0
,A_{a,s}=0,f_t(\eta^{0}_e+\eta^1_{e,x})+2\eta^1_{a,s}+A_{a,x}=0,
\end{equation}
\begin{equation}2eta^{0}_{a,s}+f\eta^0_{e,s})e^{\nu(x)}- f\eta^1_{e,t}-A_{a,t}=0 ,
2\eta^2_{a,s}+f\eta^2_{e,s})e^{\mu(x)}+f\eta^1_{e,y}+A_{a,y}=0,
\end{equation}
\begin{equation}
-\eta^1_{e,t}+(\eta^0_{a,x}+f\eta^0_{e,x})e^{\nu(x)}=0
,e^{\nu(x)}(f\eta^0_{e,y}+\eta^0_{a,y})-e^{\mu(x)}(f\eta^2_{e,t}+\eta^2_{a,t})=0,
\end{equation}
\begin{equation}e^{\nu(x)}(f\eta^0_{e,z}+\eta^0_{a,z})-e^{\mu(x)}(f\eta^3_{e,t}
+\eta^3_{a,t})=0,(f\eta^2_{e,z}+\eta^2_{a,z})+(f\eta^3_{e,y}+\eta^3_{a,y})=0,
\end{equation}
\begin{equation}
f_t\eta^0{e}+(\eta^1_{a}+f\eta^1{e})\mu(x)_x+2(\eta^0_{a,t}+f\eta^0_{e,t})-
f\xi_{e,s}-\xi_{a,s}=0,
\end{equation}
\begin{equation}
f_t\eta^0_{e}+(\eta^1_{a}+f\eta^1_{e})\mu(x)_x+2(\eta^2_{a,y}+f\eta^2_{e,y})-
f\xi_{e,s}-\xi_{a,s}=0,
\end{equation}
\begin{equation}
f_t\eta^0_{e}+(\eta^1_{a}+f\eta^1_{e})\mu(x)_x+2(\eta^3_{a,z}+f(t)\eta^3_{e,z})-
f\xi_{e,s}-\xi_{a,s}=0,\end{equation}
\begin{equation}
2(\eta^3_{a,s}+f\eta^3_{e,s})e^{\mu(x)}+f\eta^1_{e,z}+A_{a,z}=0
,2\eta^1_{a,x}-\xi_{a,s}=0,
\end{equation}
\begin{equation}
\eta^1_{a,y}+(\eta^2_{a,x}+f\eta^2_{e,x})e^{\mu(x)}=0
,\eta^1_{a,z}+(\eta^3_{a,x}+
f\eta^3_{e,x})e^{\mu(x)}=0.
\end{equation}
We solve this system by using the exact solutions of the plane symmetric static spacetime [6]. In the following section we list
all those cases where the approximate Noether symmetries(y) exist(s).
 \section{Approximate Noether symmetries from the Lagrangian of Spacetimes}
\subsection{Five Noether symmetries}
There are only five (minimal set) Noether symmetries from general form of the first order approximate Lagrangian given by eq. $(\ref{4a})$,
\begin{equation}
\textbf{Y}_0=\frac{\partial}{\partial s},\mathbf{X_0}=\frac{\partial}{\partial
t}+\frac{\epsilon s}{\alpha}\frac{\partial}{\partial s}
,\textbf{X}_1=\frac{\partial}{\partial y}
,\textbf{X}_2=\frac{\partial}{\partial z}
,\textbf{X}_3=y\frac{\partial}{\partial z}-z\frac{\partial}{\partial
y}.
\end{equation}The first integral corresponding to $X_0$ is \begin{align*}\phi_0=2[e^{\nu(x)}\dot{t}+\frac{\epsilon}{\alpha}(t\dot{t}e^{\nu(x)}-s\dot{t}^2e^{\nu(x)}+\frac{{L}s}{2})].\end{align*}
Therefore, there is only one approximate Noether symmetry, $\mathbf{X_0}$ from the Lagrangian given by eq. $(\ref{4a})$.
The Lie Algebra of these five Noether symmetries is, \\$[\mathbf{X_1},\mathbf{X_3}]=\mathbf{X_2},[\mathbf{X_2},\mathbf{X_3}]=-\mathbf{X_1},[\mathbf{X_0},\mathbf{Y_0}]=-\frac{\epsilon}{\alpha}\mathbf{Y_0}$,
$[\mathbf{X_i},\mathbf{X_j}]=0$ and $[\mathbf{X_i},\mathbf{Y_0}]=0$
otherwise.
\subsection{Six Noether Symmetries}
 We have following three cases for six Noether symmetries:\\
 (1) The first case of six Noether symmetries is
\begin{equation}\begin{split}
&ds^2=(\frac{x}{b})^2dt^2-dx^2-(\frac{x}{d})^a(dy^2+dz^2)+\\&
\frac{\epsilon t}{\alpha}\{(\frac{x}{b})^2dt^2-dx^2-(\frac{x}{d})^a(dy^2+dz^2)\},a\neq2.
\end{split}\end{equation}
The exact Noether symmetries are \begin{equation}
\textbf{Y}_0=\frac{\partial}{\partial s}
,\textbf{X}_1=\frac{\partial}{\partial y}
,\textbf{X}_2=\frac{\partial}{\partial z}
,\textbf{X}_3=y\frac{\partial}{\partial z}-z\frac{\partial}{\partial
y},
\end{equation}
\begin{equation}\textbf{Y}_1=s\frac{\partial}{\partial s }+\frac{x}{2}\frac{\partial}{\partial x}+
\frac{2-a}{4}y\frac{\partial}{\partial
y}+\frac{2-a}{4}z\frac{\partial}{\partial x},
\end{equation}
and the symmetry with approximate part is
\begin{equation}
\textbf{X}_{0}=\frac{\partial}{\partial
t}-\frac{\epsilon}{4\alpha}\{2x\frac{\partial}{\partial
x}+(2-a)y\frac{\partial}{\partial y}+(2-a)z\frac{\partial}{\partial
z}\}.\label{a}\end{equation}The conservation law or first integral corresponding to (\ref{a}) is \begin{align*}\phi_0=2[(\frac{x}{b})^2\dot{t}+\frac{\epsilon}{\alpha}(t\dot{t}(\frac{x}{b})^2+\frac{x\dot{x}}{2}+(2-a)\frac{y\dot{y}}{4}(\frac{x}{b})^a)
+(2-a)\frac{z\dot{z}}{4}(\frac{x}{b})^a].\end{align*}
$\textbf{X}_{0}$ is the approximate Noether symmetry corresponding to the energy. The Lie Algebra of these symmetry generators is\\
$[\mathbf{X_1},\mathbf{X_3}]=\mathbf{X_2},[\mathbf{X_2},\mathbf{X_3}]=-\mathbf{X_1},[\mathbf{X_1},\mathbf{Y_1}]=\frac{2-a}{4}\mathbf{X_1},$
$[\mathbf{X_2},\mathbf{Y_1}]=\frac{2-a}{4}\mathbf{X_2},[\mathbf{Y_0},\mathbf{Y_1}]=\mathbf{Y_0}$,\\
$[\mathbf{X_1},\mathbf{X_0}]=\frac{\epsilon(a-2)}{4\alpha}\mathbf{X_1},[\mathbf{X_2},\mathbf{X_0}]=\frac{\epsilon(a-2)}{4\alpha}\mathbf{X_2},
[\mathbf{X_i},\mathbf{X_j}]=0$ and $[\mathbf{X_i},\mathbf{Y_j}]=0,$
otherwise.\\
(2) The second case for six Noether symmetries is
\begin{equation}\begin{split}&ds^2=(\frac{x}{d})^bdt^2-dx^2-(\frac{x}{d})^2(dy^2+dz^2)+\\&
\frac{\epsilon t}{\alpha}\{(\frac{x}{d})^bdt^2-dx^2-(\frac{x}{d})^2(dy^2+dz^2)\}, (b\neq2).\end{split}\end{equation}
Exact Noether symmetries are
\begin{equation}
\textbf{Y}_0=\frac{\partial}{\partial s}, \textbf{X}_1=\frac{\partial}{\partial y}, \textbf{X}_2=\frac{\partial}{\partial z}, \textbf{X}_3=y\frac{\partial}{\partial z}-z\frac{\partial}{\partial y}.
\end{equation}
The symmetries with approximate parts are
\begin{equation}\textbf{Y}_{1}=s\frac{\partial}{\partial
s }+\frac{2-b}{4}t\frac{\partial}{\partial
t}+\frac{x}{2}\frac{\partial}{\partial
x}-\frac{\epsilon}{8\alpha}\{(x^2(\frac{x}{d})^{-b}+t^2\frac{(b-2)^2}{4})\frac{\partial}{\partial
t}+(2-b)xt\frac{\partial}{\partial x}\},\end{equation}
\begin{equation} \textbf{X}_{0}=\frac{\partial}{\partial
t}-\frac{\epsilon}{\alpha}\{\frac{(2-b)t}{4}\frac{\partial}{\partial
t}+\frac{x}{2}\frac{\partial}{\partial x}\}.
\end{equation} $\textbf{Y}_{1}$ and $\textbf{X}_{0}$ are the approximate Noether symmetries corresponding to scaling and energy respectively. The first integrals corresponding to $\textbf{X}_{0}$ and $\textbf{Y}_{1}$ are given in the following table
\begin{table}[H]
\begin{center}\captionof{table}{First integrals}
\begin{tabular}{|p{.5cm}|p{9cm}|}
 \hline
  Gen&First integrals\\
  \hline
$\mathbf{X_0}$&$\phi_0=2[(\frac{x}{b})^b\dot{t}+\frac{\epsilon}{\alpha}((\frac{x}{d})^b(t\dot{t}-\frac{(2-b)t\dot{t}}{4})+\frac{x\dot{x}}{2})]$\\
\hline
 $\mathbf{Y_1}$&$\phi_1=2(\frac{x}{d})^b[\frac{(2-b)t\dot{t}}{4}-2\dot{t}^2s]-x\dot{x}+2s\dot{x}^2+s{L}+$\\
 \hline
$ $&$\frac{\epsilon}{\alpha}[-\frac{x^2\dot{t}}{4}+(\frac{x}{d})^b(-\frac{7t^2\dot{t}(2-b)}{16})+2(\frac{x}{d})^bt\dot{t}^2s-
\frac{x\dot{x}(1+b)}{2}+2ts\dot{x}^2]$\\
\hline
\end{tabular}\end{center}\end{table}
  The
 Lie Algebra is\\
 $[\mathbf{X_1},\mathbf{X_3}]=\mathbf{X_2}, [\mathbf{X_2},\mathbf{X_3}]=-\mathbf{X_1},
 [\mathbf{X_0},\mathbf{Y_1}]=\frac{2-b}{4}\mathbf{X_0}$,\\
 $[\mathbf{Y_0},\mathbf{Y_1}]=\mathbf{Y_0}, [\mathbf{X_i},\mathbf{X_j}]=0$ and $[\mathbf{X_i},\mathbf{Y_j}]=0$ otherwise.\\
\subsection{Eight Noether Symmetries}
We have the following metric whose Lagrangian gives eight Noether symmetries
\begin{equation}
ds^2=(\frac{x}{b})^adt^2-dx^2-(\frac{x}{d})^a(dy^2+dz^2)+
 \frac{\epsilon t}{\alpha}\{(\frac{x}{b})^adt^2-dx^2-(\frac{x}{d})^a(dy^2+dz^2)\}.
\end{equation}
The exact Noether symmetries are given below
\begin{equation}
\textbf{Y}_0=\frac{\partial}{\partial s}, \textbf{X}_1=\frac{\partial}{\partial y}, \textbf{X}_2=\frac{\partial}{\partial z}, \textbf{X}_3=y\frac{\partial}{\partial z}-z\frac{\partial}{\partial y},
\end{equation}
and the symmetries with approximate part are
\begin{equation}
\textbf{X}_{0}=\frac{\partial}{\partial
t}-\frac{\epsilon}{8\alpha}\{2(2-a)t\frac{\partial}{\partial
t}-2x\frac{\partial}{\partial x}+2(2-a)y\frac{\partial}{\partial
y}+2(2-a)z\frac{\partial}{\partial z}\},
\label{b}\end{equation}
\begin{equation}\begin{split}\textbf{X}_{4}&=y\frac{\partial}{\partial
t}+t\frac{\partial}{\partial y}-\frac{\epsilon}{8\alpha}\{2(a-2)ty\frac{\partial}{\partial t}+xy\frac{\partial}{\partial x}\}\\&-\frac{\epsilon}{8\alpha}\{(a-2)[\frac{4x^2}{(a-2)^2}(\frac{x}{b})^{-a}-t^2-y^2+z^2]\frac{\partial}{\partial
y}+2(a-2)yz\frac{\partial}{\partial z}\},\label{c}\end{split}\end{equation}
\begin{equation}\begin{split}\textbf{X}_{5}&=z\frac{\partial}{\partial t}+t\frac{\partial}{\partial z}-\frac{\epsilon}{8\alpha}\{2(2-a)tz\frac{\partial}{\partial
t}\}\\&-\frac{\epsilon}{8\alpha}\{4xz\frac{\partial}{\partial x}-(a-2)[\frac{4x^2}{(a-2)^2}(\frac{x}{b})^{-a}-t^2+y^2-z^2]\frac{\partial}{\partial
z}+2(2-a)yz\frac{\partial}{\partial y}\},\label{d}\end{split}\end{equation}
\begin{equation}\begin{split}
\textbf{Y}_{1}&=s\frac{\partial}{\partial s}+\frac{(2-a)t}{4}\frac{\partial}{\partial
t}+\frac{x}{2}\frac{\partial}{\partial x}+ \frac{(2-a)y}{4}\frac{\partial}{\partial
y}+\frac{(2-a)z}{4}\frac{\partial}{\partial z}+\frac{\epsilon}{32\alpha}\{4x^2(\frac{x}{b})^{-a}\\&+(2-a)^2(t^2+y^2+z^2)\frac{\partial}{\partial
t}-4(a-2)tx\frac{\partial}{\partial x}+2(a-2)^2[ty\frac{\partial}{\partial y}+tz\frac{\partial}{\partial z}].
\label{e}\end{split}
\end{equation}The first integrals corresponding to the generators given in (\ref{b}), (\ref{c}), (\ref{d}) and (\ref{e}) are given in Table 2
\begin{table}[H]
\begin{center}\captionof{table}{First integrals}
\begin{tabular}{|p{.5cm}|p{13cm}|}
 \hline
  Gen&First integrals\\
  \hline
$\mathbf{X_0}$&$\phi_0=2[(\frac{x}{b})^a\dot{t}+\frac{\epsilon}{\alpha}((\frac{x}{b})^a(t\dot{t}-\frac{(2-a)t\dot{t}}{4})+\frac{x\dot{x}}{2})+
(\frac{x}{d})^a(\frac{(2-a)y\dot{y}}{4}+\frac{(2-a)z\dot{z}}{4})]$\\
\hline
 $\mathbf{X_4}$&$\phi_4=2((\frac{x}{b})^ay\dot{t}+(\frac{x}{d})^a\dot{y}t)+\frac{\epsilon}{\alpha}[(\frac{x}{b})^a\frac{(2+b)yt\dot{t}}{2}+
 (\frac{x}{d})^a\{t^2\dot{y}
 -{\dot{y}(a-2)}(\frac{x^2}{(a-2)^2}(\frac{x}{b})^{-a}-t^2-y^2+z^2)+\frac{yz^2\dot{z}(a-2)}{4}\}]$\\
 \hline
$\mathbf{X_5}$&$\phi_5=2((\frac{x}{b})^az\dot{t}+(\frac{x}{d})^a\dot{z}t)+\frac{\epsilon}{\alpha}[(\frac{x}{b})^a\frac{(2+b)zt\dot{t}}{2}+
 (\frac{x}{d})^a\{t^2\dot{z}
 -{\dot{z}(a-2)}(\frac{x^2}{(a-2)^2}(\frac{x}{b})^{-a}-t^2-y^2+z^2)+\frac{y^2z\dot{y}(a-2)}{4}\}]$\\
 \hline
$\mathbf{Y_1}$&$\phi_1=2(\frac{2-a}{2}((\frac{x}{b})^at\dot{t}-(\frac{x}{d})^ay\dot{y}-(\frac{x}{d})^az\dot{z})-
2s((\frac{x}{b})^a\dot{t}^2-(\frac{x}{d})^a\dot{y}^2-(\frac{x}{d})^a\dot{z}^2)
-x\dot{x}+2s\dot{x}^2+s{L}+\frac{\epsilon}{\alpha}[(\frac{x}{b})^a\{\frac{\dot{t}}{16}(4x^2(\frac{x}{b})^{-a}+(2-a)^2(t^2+y^2+z^2))+
2t\dot{t}(\frac{(2-a)t}{4}-s\dot{t})\}+(\frac{x}{d})^a\{-\frac{y\dot{y}t(2-a)}{4}+2ts\dot{y}^2-\frac{z\dot{z}t(2-a)}{4}+2ts\dot{z}^2\}+\frac{tx\dot{x}(a-6)}{4}+2ts\dot{x}^2]$\\
\hline
\end{tabular}\end{center}\end{table}

$\mathbf{X}_0$ corresponds to energy, $\mathbf{X}_4$, $\mathbf{X}_5$ correspond to the Lorentz transformations and $\mathbf{Y}_1$ corresponds to the scaling in the given spacetime. The Lie Algebra is\\
$[\mathbf{X_1},\mathbf{X_3}]=\mathbf{X_2},[\mathbf{X_2},\mathbf{X_3}]=-\mathbf{X_1},
[\mathbf{X_0},\mathbf{X_4}]=\mathbf{X_1},[\mathbf{X_1},\mathbf{X_4}]=\mathbf{X_0},
[\mathbf{X_3},\mathbf{X_4}]=-\mathbf{X_5},$\\$[\mathbf{X_0},\mathbf{X_5}]=\mathbf{X_2},
[\mathbf{X_2},\mathbf{X_5}]=\mathbf{X_0},[\mathbf{X_3},\mathbf{X_5}]=\mathbf{X_4},
[\mathbf{X_4},\mathbf{X_5}]=\mathbf{X_3},[\mathbf{X_0},\mathbf{Y_2}]=\frac{2-a}{4}\mathbf{X_0},$\\
$[\mathbf{X_1},\mathbf{Y_1}]=\frac{2-a}{4}\mathbf{X_1}+\frac{\epsilon(a-2)^2}{16\alpha}\mathbf{X}_{4},
[\mathbf{X_2},\mathbf{Y_1}]=\frac{2-a}{4}\mathbf{X_2}+\frac{\epsilon(a-2)^2}{16\alpha}\mathbf{X}_{5}$,\\$
[\mathbf{Y_0},\mathbf{Y_1}]=\mathbf{Y_0},[\mathbf{X_i},\mathbf{X_j}]=0$,$[\mathbf{X_i},\mathbf{Y_j}]=0$ and $[\mathbf{Y_i},\mathbf{Y_j}]=0$ otherwise.\\

\subsection{Nine Noether Symmetries}
The Lagrangian of following metric gives nine Noether symmetries in which four have approximate part
\begin{equation}ds^2=(\frac{x}{a})^2dt^2-dx^2-(\frac{x}{b})^2(dy^2+dz^2)+
\frac{\epsilon t}{\alpha}\{(\frac{x}{a})^2dt^2-dx^2-(\frac{x}{b})^2(dy^2+dz^2)\}.
\end{equation}
Exact symmetries are
\begin{equation}
\textbf{Y}_0=\frac{\partial}{\partial s}, \textbf{X}_1=\frac{\partial}{\partial y}, \textbf{X}_2=\frac{\partial}{\partial z},
\end{equation}
\begin{equation}\textbf{X}_3=y\frac{\partial}{\partial z}-z\frac{\partial}{\partial y}, \textbf{Y}_{1}=2s\frac{\partial}{\partial s}+x\frac{\partial}{\partial x},
\end{equation}
and the approximate Noether symmetries are
\begin{equation}
\textbf{X}_{0}=\frac{\partial}{\partial t}-\frac{\epsilon}{2\alpha}x\frac{\partial}{\partial
x}, \textbf{Y}_{2}=s^2\frac{\partial}{\partial s}+xs\frac{\partial}{\partial
x}-\frac{\epsilon b^2}{2\alpha}s\frac{\partial}{\partial t},
\end{equation}
\begin{equation}\textbf{X}_{4}=y\frac{\partial}{\partial t}+t\frac{\partial}{\partial y}-\frac{\epsilon}{2\alpha}\{xy\frac{\partial}{\partial
x}-b^2ln(\frac{x}{k})\frac{\partial}{\partial y}\},\end{equation}
\begin{equation}
\textbf{X}_{5}=z\frac{\partial}{\partial t}+t\frac{\partial}{\partial z}-\frac{\epsilon}{2\alpha}\{xz\frac{\partial}{\partial
x}-b^2ln(\frac{x}{k})\frac{\partial}{\partial z}\}.
\end{equation}The first integrals corresponding to $\mathbf{X}_0$, $\mathbf{Y}_2$, $\mathbf{X}_4$ and $\mathbf{X}_5$ are given in the following Table
\begin{table}[H]
\begin{center}\captionof{table}{First integrals}
\begin{tabular}{|p{.5cm}|p{12cm}|}
 \hline
  Gen&First integrals\\
  \hline
$\mathbf{X_0}$&$\phi_0=2[(\frac{x}{a})^2\dot{t}+\frac{\epsilon}{\alpha}\frac{x\dot{x}}{2}]$\\
\hline
 $\mathbf{Y_2}$&$\phi_2=-2(\frac{x}{a})^2s\dot{t}-2x\dot{x}s+2s^2\dot{x}^2+x^2-
 \frac{\epsilon}{\alpha}[(\frac{x}{a})^2(a^2s\dot{t}+2st\dot{t}^2)+2t\dot{x}(xs-s^2x)+x^2t]$\\
 \hline
$\mathbf{X_4}$&$\phi_4=2((\frac{x}{a})^2y\dot{t}-(\frac{x}{b})^2\dot{y}t)+
\frac{\epsilon}{\alpha}[2(\frac{x}{a})^2t\dot{t}y-(\frac{x}{b})^2\dot{y}b^2\ln\frac{x}{k}-2(\frac{x}{b})^2\dot{y}t^2)+x\dot{x}y]$\\
\hline
$\mathbf{X_5}$&$\phi_5=2((\frac{x}{a})^2z\dot{t}-(\frac{x}{b})^2\dot{z}t)+
\frac{\epsilon}{\alpha}[2(\frac{x}{a})^2t\dot{t}z-(\frac{x}{b})^2\dot{z}b^2\ln\frac{x}{k}-2(\frac{x}{b})^2\dot{z}t^2)+x\dot{x}z]$\\
\hline
\end{tabular}\end{center}\end{table}

$\mathbf{X}_0$ corresponds to the energy and $\mathbf{X}_4$ and
$\mathbf{X}_5$ correspond to the Lorentz transformations.\\
The Lie algebra is
\\$[\mathbf{X_1},\mathbf{X_3}]=\mathbf{X_2},[\mathbf{X_2},\mathbf{X_3}]=-\mathbf{X_1},[\mathbf{X_0},\mathbf{X_4}]=\mathbf{X_1},
[\mathbf{X_3},\mathbf{X_4}]=-\mathbf{X_5},[\mathbf{X_0},\mathbf{X_5}]=\mathbf{X_2},$\\$[\mathbf{X_3},\mathbf{X_5}]=\mathbf{X_4},
[\mathbf{X_4},\mathbf{X_5}]=\mathbf{X_3},[\mathbf{Y_0},\mathbf{Y_2}]=\mathbf{Y_1}-\epsilon\frac{b^2}{4\alpha}\mathbf{X_0},[\mathbf{Y_2},
\mathbf{\mathbf{Y_2}}]=\mathbf{\mathbf{Y_2}},[\mathbf{X_i},\mathbf{X_j}]=0,$\\$[\mathbf{X_1},\mathbf{X_4}]=\mathbf{X_0},
[\mathbf{Y_1},\mathbf{Y_2}]=2\mathbf{Y_2},[\mathbf{X_2},\mathbf{X_5}]=\mathbf{X_0},[\mathbf{Y_1},\mathbf{X_4}]=\epsilon\frac{b^2}{2\alpha}\mathbf{X_1}
,[\mathbf{Y_1},\mathbf{X_5}]=\epsilon\frac{b^2}{2\alpha}\mathbf{X_2},$\\
$[\mathbf{X_0},\mathbf{X_4}]=\mathbf{X_1},[\mathbf{X_0},\mathbf{X_5}]=\mathbf{X_2},
[\mathbf{Y_2},\mathbf{X_4}]=\epsilon\frac{b^2}{2\alpha}s\frac{\partial}{\partial
y},
[\mathbf{Y_2},\mathbf{X_5}]=\epsilon\frac{b^2}{2\alpha}s\frac{\partial}{\partial
z},$\\ $
[\mathbf{X_4},\mathbf{X_5}]=\mathbf{X_3},[\mathbf{X_i},\mathbf{Y_j}]=0$
and $[\mathbf{Y_i},\mathbf{Y_j}]=0$ otherwise.
\section{Observations}
After getting a complete list of possible (by taking time dependent approximate factor) plane symmetric spacetimes where approximate Noether's symmetries exist we have the following observations:\\
(i) Approximate part appears in cases of  five, six, eight or nine Noether symmetries for the Lagrangian of the given plane symmetric spacetimes.\\
(ii) Only time translational, scaling and the symmetry corresponds to Lorentz transformation admit approximate parts that is exact energy conservation, exact scaling and exact Lorentz transformation lost, there are some approximation to all these three quantities .\\
(iii) Linear momentum, angular momentum and quantities corresponding to  Galilean transformation are conserved for plane symmetric spacetimes as there is no approximate Noether symmetry corresponding to these qauntities.\\
(iv) The Lagrangian of spacetimes with a section of zero curvature [9] do not give approximate Noether's symmetries. This means that when a section of zero curvature appears in the spacetimes the approximate symmetries disappear.\\
(v) Approximate Noether symmetry does not exist in flat spacetimes (Minkowski spacetime).\\

 It is well known that $KV\subseteq HV\subset NS$, where KV stand for the Killing vectors, HV for homothetic vectors and NS for Noether symmetries. A such relation for the approximate symmetries has not been seen in the literature. However, on the basis of our results it can be conjectured that a similar result would hold for approximate symmetries as well. It will be interesting to classify spacetimes by their approximate KV and HV to prove the conjecture.

\end{document}